\documentclass[letter, conference]{IEEEtran}

\IEEEoverridecommandlockouts
\usepackage{cite}
\usepackage{amsmath,amssymb,amsfonts}
\usepackage{algorithmic}
\usepackage{graphicx}
\usepackage{textcomp}
\usepackage{xcolor}
\usepackage{subfigure}
\def\BibTeX{{\rm B\kern-.05em{\sc i\kern-.025em b}\kern-.08em
    T\kern-.1667em\lower.7ex\hbox{E}\kern-.125emX}}
    
\usepackage{soul}

\begin{document}

\title{\LARGE \bf
DLT federation for Edge robotics}

\author{\IEEEauthorblockN{Kiril Antevski\IEEEauthorrefmark{1}, 
Milan Groshev\IEEEauthorrefmark{1}, Gabriele Baldoni\IEEEauthorrefmark{2}, Carlos J. Bernardos\IEEEauthorrefmark{1}
}\\
\IEEEauthorblockA{
\IEEEauthorrefmark{1}University Carlos III of Madrid, Spain,
\IEEEauthorrefmark{2}ADLINK Technology, France
}
\\[-6.5ex]
}

\maketitle

\begin{abstract}
The concept of federation in 5G and NFV networks aims to provide orchestration of services across multiple administrative domains. Edge robotics, as a field of robotics, implements the robot control on the network edge by relying on low-latency and reliable access connectivity.
In this paper, we propose a solution that enables Edge robotics service to expand its service footprint or access coverage over multiple administrative domains. 
We propose application of Distributed ledger technologies (DLTs) for the federation procedures to enable private, secure and trusty interactions between undisclosed administrative domains. 
The solution is applied on a real-case Edge robotics experimental scenario. 
The results show that it takes around 19 seconds to deploy \& federate a Edge robotics service in an external/anonymous domain without any service down-time.

\end{abstract}


\section{Introduction}
\label{sec:introduction}




In recent years, Edge robotics emerged as a consequence of the rapid development of Edge computing to address the network performance (e.g., high latency, unpredictable jitter) related challenges that Cloud-based robotic applications experience\cite{cloudrobotics}. By placing computing and storage resource near the edge, robotic systems can execute applications closer to the robots resulting in more predictable communication and overall better system performance.
For the market, the Edge robotics services are an opportunity for mobile robots to be employed in accomplishing a range of manual tasks (e.g., security/surveillance, cleaning, delivery of goods, collecting fruits, ehealth emergency response, sports video coverage, etc.). 
The linchpin of the Edge robotics is the constant robots connectivity over the access network and the available real-time information about the connectivity. 
The high mobility of robots demands change of the point of access in the access network
which is currently feasible within a single administrative domain.
Often enough, an Edge robotics service require fast and short-lasting expansion of the service footprint over multiple administrative domains (e.g., delivery of goods for a big day-lasting events, emergency response for large area, video streaming of cycling events, etc.).


With the advancement of network function virtualization (NFV) and Multi-access Edge Computing (MEC) in the 5G networks, the concept of \emph{federation} has been introduced. The federation as a 5G networks concept, enables orchestration of resources and services across multiple administrative domains. Virtualized access networks enable the robotic service providers to request on-demand deployment of virtualized access point, in an external administrative domain at a specific location through the federation process.
With the introduction of the Fog concept, where volatile and low-power consumption devices are used as access network, federation extends the eco-system heterogeneity and variance in the access network coverage. Multiple administrative domains can simultaneously deploy virtualized wireless networks over range of hardware devices thanks to Fog, Edge, MEC and NFV concepts. 

As a consequence, a higher number of involved administrative domains, eligible to provide on-demand federation of services and resources, increase the risk of security threats, maintaining SLAs, privacy violations, and etc. The Distributed Ledger Technology (DLT) is a potential solution to counter the negative byproducts of the federation process. The Blockchain as a DLT, by default provides trust, security and cryptography to participants. Leveraging the DLTs, the administrative domain can discover, negotiate and federate services on-the-fly. 

In this paper, our goal is to (\emph{i}) propose a DLT federation concept in the Edge robotics environment, (\emph{ii}) apply the DLT federation concept on a real Edge robotics test-bed and (\emph{iii}) evaluate the performance of the solution. 

The rest of the paper is organized as follows. Sec.~\ref{sec:background} provides background on the federation concept, DLT and related work. Sec.~\ref{sec:concept} explains a MECinNFV-based Edge robotics service, the federation procedures and how DLT fits into it. Sec.~\ref{sec:experiments} describes the experimental scenario and testbed, while  Sec.~\ref{sec:results} detail the performed experiments and summarizes the obtained results. Finally, we conclude the paper in Sec.~\ref{sec:conclusion}.

\section{Background}
\label{sec:background}




\subsection{Federation in NFV}
\label{subsec:background_federation}

In a NFV environment, the federation mechanism is used by administrative domains (ADs) to deploy network services or allocate resources over the infrastructure of an external domain. 
The federation procedure is triggered by a consumer domain as a need of extending specific service (e.g., at specific geo-location) or the lack of constituent resources. 

Federation requires all collaborative administrative domains to have mutual cooperation agreements. Typically, these agreements are signed by business executives. Upon agreement, the agreed ADs trust each other and establish a peer-to-peer connectivity among themselves or define a trusted centralized entity to manage their interactions.
Therefore, federation interactions can be executed in a centralized, decentralized or distributed manner.  
In a centralized solution all involved ADs have to sign mutual agreements and trust a centralized entity located in a neutral location. The centralized entity is managed by the joint community of involved ADs and oversees the federation interaction, acting as neutral "middle-man". The positive side is that it is a highly trusty and scalable solution, however the disadvantage is the set-up time, joint effort, continuous maintenance and that it is a single-point of failure system.  
A decentralized solution is the simplest to employ, at the cost of the lowest scalability. Each AD establishes peer-to-peer connectivity with each external AD. 
This implies that each new connection is followed by business meeting and connectivity set-up (e.g., 50 connections would take at least 50 or more days). 
Several approaches rely on this type of federation~\cite{so_5gt_orch_fed}~\cite{5gt_demo}. 
A distributed solution is a hybrid approach, more similar to the centralized solution, where the central entity is distributed in each AD. Through application of distributed ledger technologies (DLTs), the joint set-up effort is lower than the centralized solution, while the benefits are the same without a single-point of failure. Applying a Blockchain as a DLT increases the privacy, trust and security~\cite{kiril_2}. 

\subsection{Blockchain \& Smart-contracts overview}
\label{subsec:blockchain_overview}

The \emph{blockchain} technology originated as a driving mechanism for Bitcoin~\cite{bitcoin_nakamoto}, providing a distributed and secure ledger that records every transaction between anonymous users. 
The blockchain itself, contains distributed time-stamped blocks filled with transactions that can contain any data. Each block points to the prior block, creating a chain of blocks in history until the genesis block (\emph{block 0}). The blocks are generated by nodes (e.g., any computing device) interconnected in a peer-to-peer blockchain network. 

The goal is to maintain a single block creation at time for the whole blockchain network. Thus the nodes run consensus mechanism to validate the single block creation process. Various consensus mechanisms exist from more simple (Byzantine Fault Tolerance - BFT, Proof-of-Authority - PoA), to more computational expensive such as Proof-of-Work (PoW) (used in Bitcoin), or incentive based such as Proof-of-Stake (PoS)~\cite{consensus2_survey}.
The consensus mechanism has the role of maintaining the blockchain distributed, secure, trust and privacy features,  without the need of 3rd-party centralized entity.

There are two types of blockchain networks: permissionless and permissioned. 
Permissionless blockchain networks are open and referred to as public blockchain networks. Most popular persmissionless blockchain networks are Bitcoin, Ethereum~\cite{ethereum_main}, etc.
Permissioned blockchain networks are private networks where the participating nodes and the users are known to each other or belong to a central organization, group or etc. Most popular permissioned blockchains are Hyperledger~\footnote{https://www.hyperledger.org/}
and Corda~\footnote{https://www.corda.net/}

Ethereum, adapts a concept of \emph{Smart Contracts}. 
The smart contract is a set of binary code, similar to a computer application, that runs on top of blockchain. Once the smart contract is deployed on the blockchain, it is immutable and operates independently (from its creator) with its own blockchain address. 
Users send transactions with input data to the smart contract, which in turn executes itself (bytecode) to produce an output data, that can be permanently stored on the blockchain.

With the adoption of smart contracts, the application of the blockchain technology significantly expands in solving lots of general problems that require a middle-man in the process.

\subsection{Related works in Edge robotics}
\label{subsec:related_works}

Following the principles of Edge robotics, \cite{edge_robotics_1}~elaborates on the edge-computing friendly functionalities in healthcare robots and discusses the corespondent edge computing techniques in order to materialize wireless driven healthcare robotic services. Moreover, an example of system architecture that exploits the edge to achieve offloading for computationally
expensive localization and mapping is presented in~\cite{edge_robotics_2}. In~\cite{edge_robotics_3}, the authors present the possibilities of deploying AI based dynamic robotic control in the edge of the network to self-balance service robot and pick up a box automatically. The experimental test-bed and scenario (described in the following sections) is partially implemented in our previous work~\cite{conext}. 

\section{Federation in Edge robotics}
\label{sec:concept}


In this section, first we dive into the Edge robotics service and realization through Multi-access Edge Computing (MEC) in NFV. Then, as a consequence of the dynamic and volatile environment, we propose the edge federation concept. Finally, we explain how DLT can be applied for private, secure, and trusty edge federation.  

\subsection{Edge robotics: MECinNFV-based service}
\label{subsec:building_blocks}

Driven by the opportunities of being in close proximity to the end-user, ETSI created the Multi-access Edge Computing (MEC) framework as one of the early implementations of Edge computing. In this work, the Edge robotics service relies on the MEC in NFV reference architecture~\cite{etsi.mec.017}~\cite{mec_in_nfv_kiril}.
In this realization of the MEC, the key components of the architecture (e.g., MEC platform, MEC applications, and MEC services) are realized as virtualized network functions (VNFs) over a virtualized infrastructure.
To that end, the Edge robotics service is represented by MEC apps distributed between robots and MEC hosts. The points of access (e.g., virtual access point) are represented by a MEC apps as well,
while MEC services provide real-time radio network information or robot localization information through a MEC platform. The Edge robotic service can use this information to dynamically adapt the robot operations. 

\begin{figure}
	\centering
	\includegraphics[width=6cm]{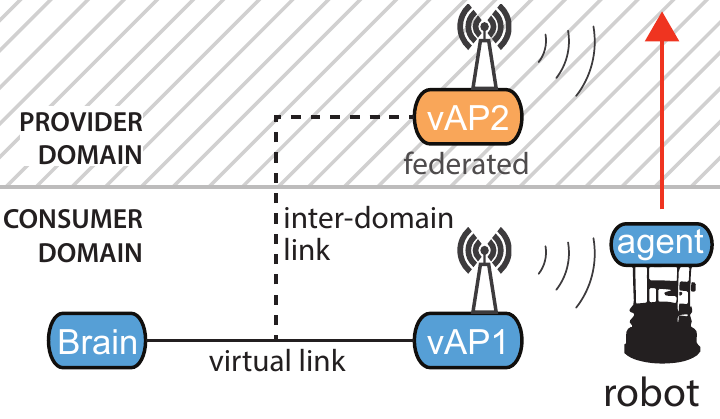}
	\caption{Edge service}
	\vspace{-5 mm}
	\label{fig:edge_service}
\end{figure}

The combination of radio context information and location coordinates allow  
the robot to move within the boundaries of a single administrative domain.
Our proposal is that through application of service federation, an Edge robotics service would not be limited (to single AD) and it would be able to extend the desired service footprint at anytime, anywhere.  
A simplified service federation of an Edge robotics service is illustrated on Fig.~\ref{fig:edge_service}. 
All colored blocks represent MEC apps as VNFs. The blue blocks present the MEC apps of an exemplary Edge robotics service deployed in a consumer domain. The "Brain" contains the control logic that provides movement instructions to a robot "agent", through the virtual access point ("vAP1"). The robot, via the "agent", executes the movement commands using its actuators and provides real-time sensor data back to the "Brain". In short, this is a closed-loop which allows the "Brain" to control the robot to accomplish different tasks (for more details refer to~\cite{conext}).
When the robot leaves the coverage area of the consumer domain, a service federation is initiated by the consumer domain. The service federation includes deployment of new virtual access point ("vAP2") in a provider domain that can ensure extended network coverage. The federated "vAP2" establishes an overlay connection to the "Brain" through an inter-domain link. Once the end-to-end connectivity is established, the closed-loop between the "Brain" and the robot continues through the federated "vAP2" without any service interruption.

\subsection{Service federation procedures}
\label{subsec:edge_federation}

In our solution we focus on the service federation rather than resource federation. In service federation a consumer domain (orchestrator) requests an extension of a service (or part of a service) to be deployed over a provider domain. The provider domain (orchestrator) oversees the complete deployment process of the service extension. While in resource federation the provider domain only provides available resources (e.g., computing or networking) to the consumer domain, and the deployment of the service extension is executed by the consumer domain. 
In order to successfully complete a service federation~\cite{jordi_vtmag}~\cite{5gt_infocom}, there are several federation procedures that are executed in sequence:

\begin{itemize}
	\item \textbf{Registration} - initial procedure through which the administrative domains establish their peer-to-peer inter-connectivity or register to a central entity. The registration procedure characterizes the type of federation, which can be relatively open or strictly closed. As an open federation can be considered when external new domains can more easily register to the peer-to-peer or centralized interaction. 
	The closed federation includes pre-defined participants with strict policies and rules, manually set and defined by the ADs. 
	\item \textbf{Discovery} - in this procedure the participating ADs periodically broadcast or exchange among themselves information on their capabilities to provide services or (computing/networking) resources. 
	Each AD creates and continuously updates a global view of the available service or resources at the external ADs.
	\item \textbf{Announcement} - this procedure is triggered by the consumer domain, once it has been decided the need to federate part of a service in an external peering domain. An announcement is broadcast to all potential provider ADs. The announcement conveys the requirements for a given service or set of resources. In the centralized case, the central entity is used as a proxy.  
	\item \textbf{Negotiation} - the potential provider ADs receive the announced offer, analyze if they can satisfy the requirements and send back a positive or negative answer. The positive answer includes the pricing of the service.
	\item \textbf{Acceptance \& deployment} - the consumer AD analyzes all collected answers and chooses an offer of a single provider domain. The selection process is entirely left to the consumer AD's internal policies and preferences. The consumer domain sends an acceptance reply to the chosen provider AD. The provider AD starts the deployment of the requested \emph{federated} service.
	\item \textbf{Usage \& Charging} - once the provider AD deploys the federated service, it notifies the consumer AD and sends all necessary information for the consumer AD to include the federated service as part of the end-to-end service deployment. From that on, the provider AD starts charging for the federated service during its life-cycle, until it is terminated.
\end{itemize}

Please note that the security/privacy and trust among the participating ADs is vital in all the aforementioned procedures. Actually, due to competitive reasons, any AD (e.g., mobile operators, cloud providers, etc.,) would not reveal much information regarding the underlying internal infrastructure or the full capabilities for service deployments.

\subsection{Applying DLT for federation}
\label{subsec:dlt_application}

Depending on how the service federation procedures (described in Sec.~\ref{subsec:edge_federation}) are realized, the sequential completion of the whole federation process can take more than a minute or even an hour. In a dynamic and heterogeneous environment, where the underlying infrastructure of each domain is continuously modified, the state can change in order of seconds. 

To boost the federation process in secure manner, our idea is to squeeze the whole service federation process (from Sec.~\ref{subsec:edge_federation}) to run on a DLT.
More specifically, the federation procedures to be stored and deployed on a Federation smart-contract (SC) which is running on top of a permissioned blockchain. 
The design of the Federation SC is completely open. Our focus in the smart-contract design is to maintain neutrality and privacy while overseeing the federation procedures that involve all ADs. 

Each domain sets up a single node as part of the peer-to-peer blockchain network.
The distributed nature of the blockchain network allows scalability while maintaining the security. 
As mentioned in Sec.~\ref{subsec:blockchain_overview}, the ADs communicate with the Federation SC through transactions. The transactions are recorded in the blocks. The sealing or generation of blocks depends on the consensus protocol. The choice of the consensus protocol would determine the speed and the security level of the federation process. 
For example, the \emph{Proof-of-Authority} (PoA) consensus increases the speed, while the \emph{Proof-of-Work} (PoW) mechanism increases the security of the blockchain.

Each new joining AD establishes connectivity with at least a single node in the blockchain network using a new and locally deployed node. Then, it registers to the Federation SC with a single-transaction registration using its unique blockchain address.
In the single-transaction registration the Federation SC records the information of the registering AD and its service footprint. This way the registration procedure explained in Sec.~\ref{subsec:edge_federation} is relatively simple to be realized.
Once the registration procedure is successfully completed, the AD is ready to consume or provide federated services.

\begin{figure}
	\centering
	\includegraphics[width=7.3cm]{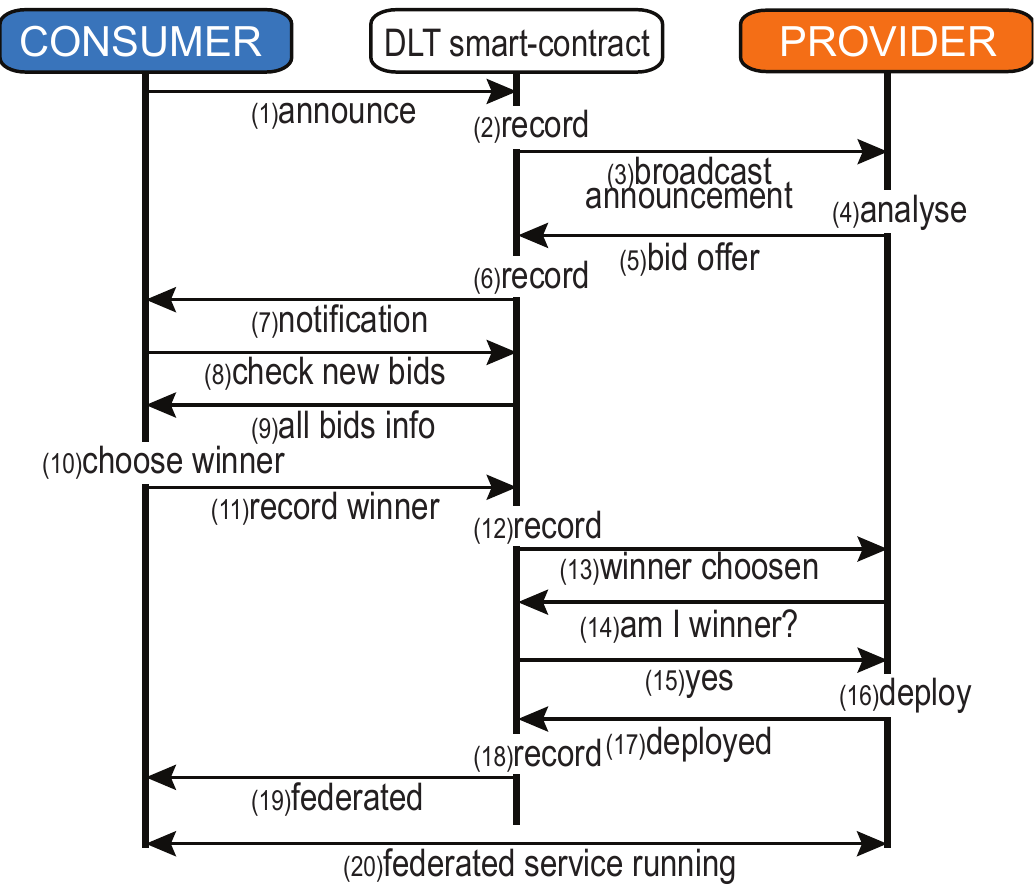}
	\caption{Sequence message diagram for Federation Smart-Contract and administrative domains during federation}
	\vspace{-5 mm}
	\label{fig:federation_sc}
\end{figure}

Fig.~\ref{fig:federation_sc} presents the interactions of registered ADs with the Federation SC for a single service federation process. 
The registered ADs can participate as consumers or providers in the federation process. 
When a consumer AD needs a federated services, it creates a federation \textbf{announcement} (step 1).
The announcement is sent as a transaction to the Federation SC which records the announcement as a new auction process on the blockchain (step 2). 
Then, the Federation SC broadcasts the auction to all registered ADs (step 3). 
Note that the address of the consumer AD is hidden in the broadcast announcement in order to protect the AD's privacy and prevent the rest of the ADs to passively collect information. 
Thus, the \textbf{discovery} phase is omitted in the design of the Federation SC. 
Instead, our approach is using a single-blinded reverse auction~\cite{reverse_auction}, where a consumer AD anonymously creates an announcement offer and the rest of the potential provider ADs are bidding for it.
Therefore, once the broadcast announcement is received, the potential providers analyze the requirements and place a bid offer to the Federation SC (step 4 \& 5). Each received offer is mapped and recorded by the Federation SC (step 6). 

In our vision the Federation SC is used more as a tool for maintaining neutrality and privacy than a governing or an authority member in the federation process. As a result, the bidding process is controlled by the consumer AD. That way the consumer AD has the full control and freedom to apply any selection policies (e.g., prioritize given offers, select the lowest price offer, etc.). In other words, the consumer domain is oversees the \textbf{negotiation} and \textbf{acceptance} procedures.
Therefore, the consumer AD is notified for any new bidding offer and it polls the Federation SC to obtain the information of each bidding offer (step 7, 8 \& 9). Once the consumer AD selects a provider AD (e.g., winning provider), it closes the auction in the Federation SC (step 10 \& 11). 
The winning provider is recorded by the Federation SC, which immediately broadcasts message to all participating ADs that the federation announcement has finished and a winner is chosen (step 12 \& 13).
Each of the participating ADs attempts to obtain the details in order to deploy the federated service. As shown on Fig.~\ref{fig:federation_sc}, only the winning provider AD has the granted access to the information (step 14 \& 15). 

The information that the provider domain obtains can vary depending of the trust-level of the participating ADs. If the participating ADs want to maintain privacy, having low trust level towards other ADs or \textbf{untrusty communication}, the service deployment information is limited (e.g., descriptor to be used, consumer's endpoint to establish data-connectivity).
If the participating ADs have higher level of trusts or \textbf{trusty communication}, the information that the provider domain can access is broader (e.g., database of resources, storage, different endpoints, etc.). 
At this point the negotiation and acceptance phases (of Sec.~\ref{subsec:edge_federation}) are completed and the \textbf{deployment} of the federation service has started (step 16). 


Once the deployment is concluded, the provider AD confirms the operation by sending transaction to the Federation SC (step 17). The Federation SC records the successful deployment and initiates charging for the federated service (step 18). The \textbf{charging} can be applied through micropayment channels~\cite{payment_ch_ethereum}. The micropayment channel applied on the blockchain can enable single non-bias charging record that is immutable for both the consumer AD and provider AD. 

At the end, the Federation SC notifies the consumer AD of successful federated service deployment (step 19 \& 20). The consumer AD leverages the running federated service until is needed, then terminates the service through the Federation SC. The termination procedure is omitted in this work. 

\section{Experimental setup}
\label{sec:experiments}

To prove the feasibility of the DLT federation for Edge robotics (of Sec.~\ref{sec:concept}) and evaluate the solution, we have deployed an experimental test-bed which on top of it we run trusty \& untrusty experimental scenario. 

\begin{figure}
	\centering
	\vspace{3 mm}
	\includegraphics[width=\columnwidth]{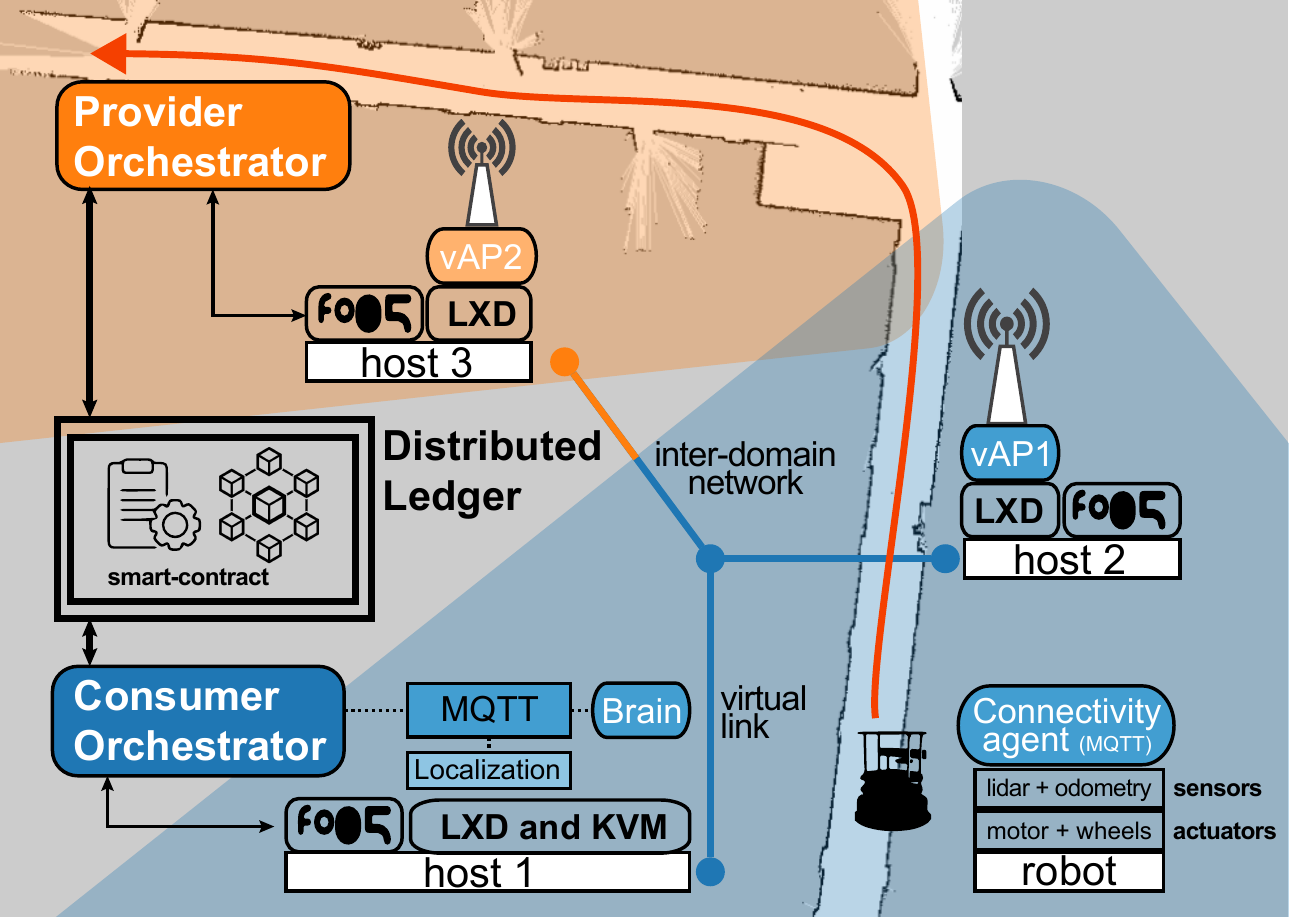}
	\caption{Edge robotics experimental test-bed \& scenario}
	\vspace{-5 mm}
	\label{fig:scenario}
\end{figure}

The experimental test-bed (shown on Fig.~\ref{fig:scenario}) consists of a robot, Ethereum blockchain node and two administrative domains (ADs) - consumer and provider domain - with their underlying infrastructure. The test-bed is deployed along a hallway in the University Carlos III of Madrid.

The consumer domain infrastructure consists of two MEC hosts depicted as host 1 and host 2 on Fig.~\ref{fig:scenario}. KVM and LXD virtualization is running on top of host 1, while only LXD on top of host 2. Both hosts are orchestrated by the Consumer orchestrator which in this case is a simple custom developed orchetrator determined for the whole scenario process. An on-boarded Edge robotics service is similar to the service described in Fig.~\ref{fig:edge_service}. The Consumer orchestrator deploys the Edge robotics service over the underlying infrastructure (host 1 \& host 2) through the distributed Virtualized Infrastructure Manager (VIM) - Fog05\footnote{https://fog05.io/}. 
As described in Sec.~\ref{subsec:edge_federation}, the Edge robotics service is deployed as VNF-MEC apps (shown as blue rounded boxes on Fig.~\ref{fig:scenario}):
\begin{itemize}
	\item \emph{Brain} is a MEC app deployed over host 1.
	\item \emph{vAP1} is deployed over host 2 as hostapd MEC app, inter-connected through virtual link to Brain.
	\item \emph{Connectivity agent} is deployed over the robot hardware. The robot hardware consists of motor wheels as actuators, 802.11 connectivity, and sensors (lidar \& odometry).
\end{itemize}
A MQTT broker is substituting the role of a MEC platform. The \emph{Brain}, as a main MEC application, is consuming a Localization MEC service via the MQTT broker. 

The provider domain is isolated from the consumer domain. Contains a single host (illustrated as host 3 on Fig.~\ref{fig:scenario}). The Provider orchestrator is a replica of the consumer orchestrator that orchestrates the virtualized infrastructure (LXD) through new instance of Fog05. The provider orchestrator has only the on-boarded image of the \emph{vAP2} MEC application on Fog05 (illustrated as orange rounded box on Fig.~\ref{fig:scenario}). 

The Distributed Ledger contains two instances of Ethereum blockchain. The instances are deployed over a virtual machine on a server at the University network. Both instances contain the Federation SC described in Sec.~\ref{subsec:dlt_application}. The first instance is running \emph{Proof-of-Authority} (PoA) consensus for trusty communication, and the second instance \emph{Proof-of-Work} (PoW) for untrusty communication.
In-depth consensus mechanism comparison is out of scope for this work.

The experimental scenario is mimicking a real use-case where the robot is instructed to deliver goods or clean an area at the University, following a path as illustrated with the red line on Fig.~\ref{fig:scenario}. In order to finalize the task,
the robot needs to drive from the blue (consumer) domain to the area of coverage of the orange (provider) domain. The \emph{Brain} is aware of the real-time robot's location by consuming the Localization MEC Service. The \emph{Brain} triggers the federation procedure to the consumer orchestrator when the robot approaches the boundaries of the vAP1 coverage. 
On triggering event, the consumer orchestrator proceeds with the federation procedure as described in Sec.~\ref{subsec:dlt_application} and Fig.~\ref{fig:federation_sc}. The provider domain, as a winner, establishes an overlay inter-domain link to the consumer domain, and deploys the \emph{vAP2} (as depicted on Fig.~\ref{fig:scenario}). 
After the deployment of the federated \emph{vAP2} has finished, the provider orchestrator confirms the deployment to the Federation SC by storing the BSSID of the deployed \emph{vAP2}. The consumer domain delivers this information to the \emph{Brain}. Finally, the \emph{Brain} instructs the robot, or the \emph{Connectivity agent}, to switch connectivity to 
the BSSID of \emph{vAP2}. The \emph{Connectivity agent} connects to the \emph{vAP2} while the closed-loop (\emph{Brain} to robot) is not broken. The closed-loop data in both directions starts passing through the overlay inter-domain link.

\section{Results}
\label{sec:results}


In this section we are evaluating the time performance of the Edge robotics federation using DLT by running the experimental scenario as described in Sec.~\ref{sec:experiments}. We run the experimental scenario using (\emph{i}) the PoA-based blockchain instance that uses trusty communication between the domains, and (\emph{ii}) the experimental scenario with the default PoW-based consensus and untrusty communication. As already mentioned, in the untrusty communication the consumer provides only the inter-domain link endpoint, and the provider domain provides back the BSSID 
of the \emph{vAP2}, upon deployment. We made a number of experimental runs for each of the PoA-based and PoW-based scenarios. In the rest of the section we present the average times for each step in the process.

\begin{figure}
	\begin{subfigure}
		\centering
		\includegraphics[width=\columnwidth]{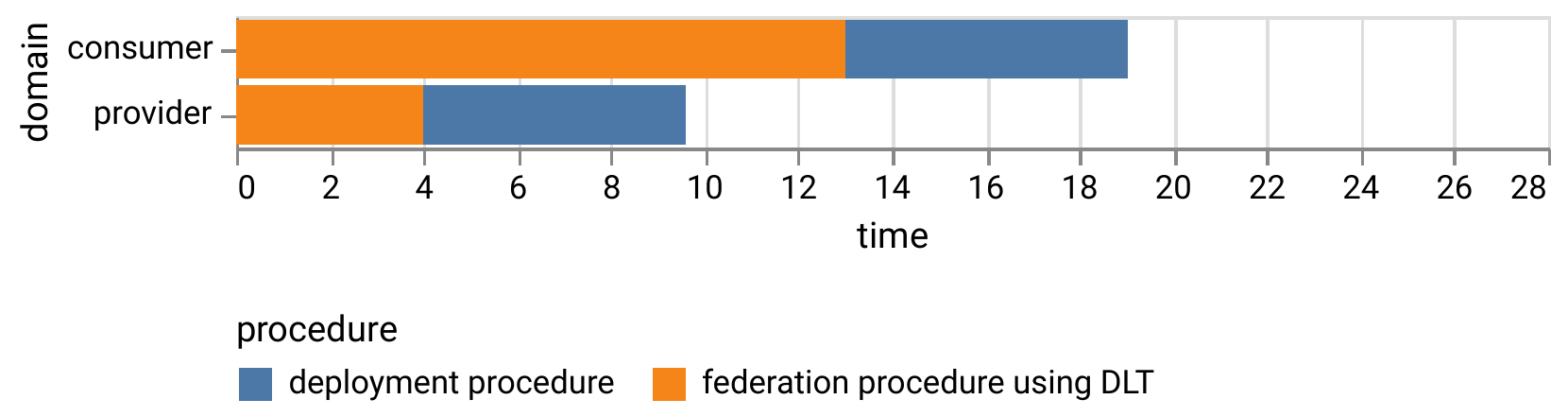}
	\end{subfigure}
	\begin{subfigure}
		\centering
		\includegraphics[width=\columnwidth]{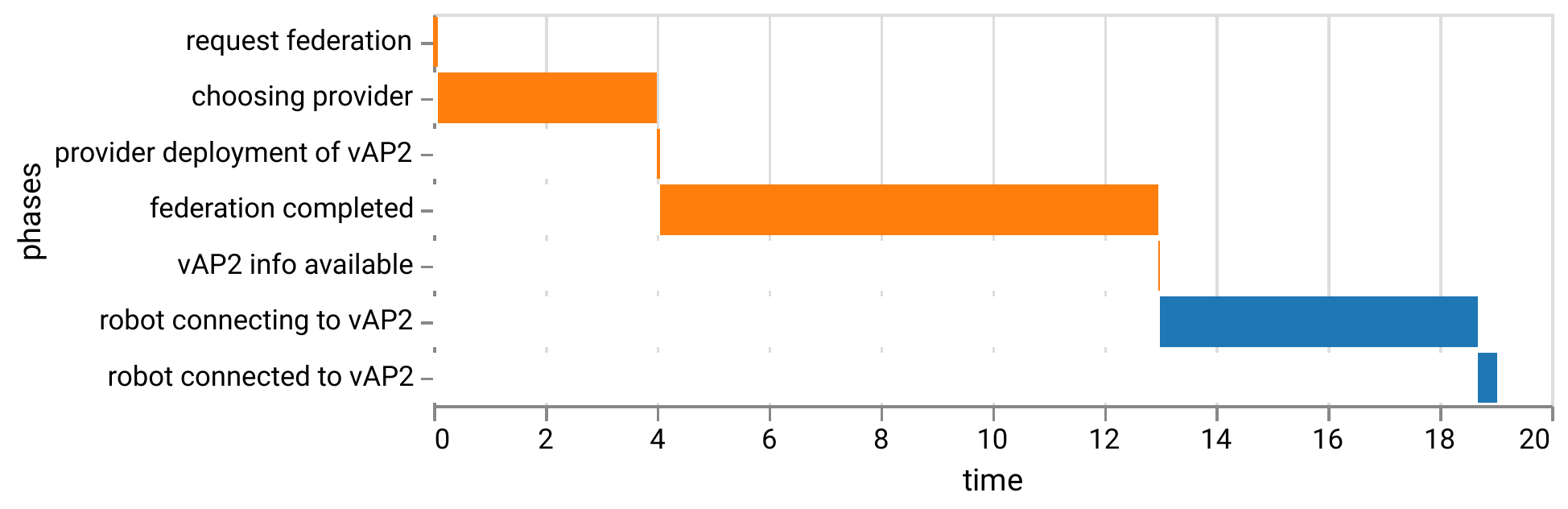}
	\end{subfigure}
	
	\begin{subfigure}
		\centering
		\includegraphics[width=\columnwidth]{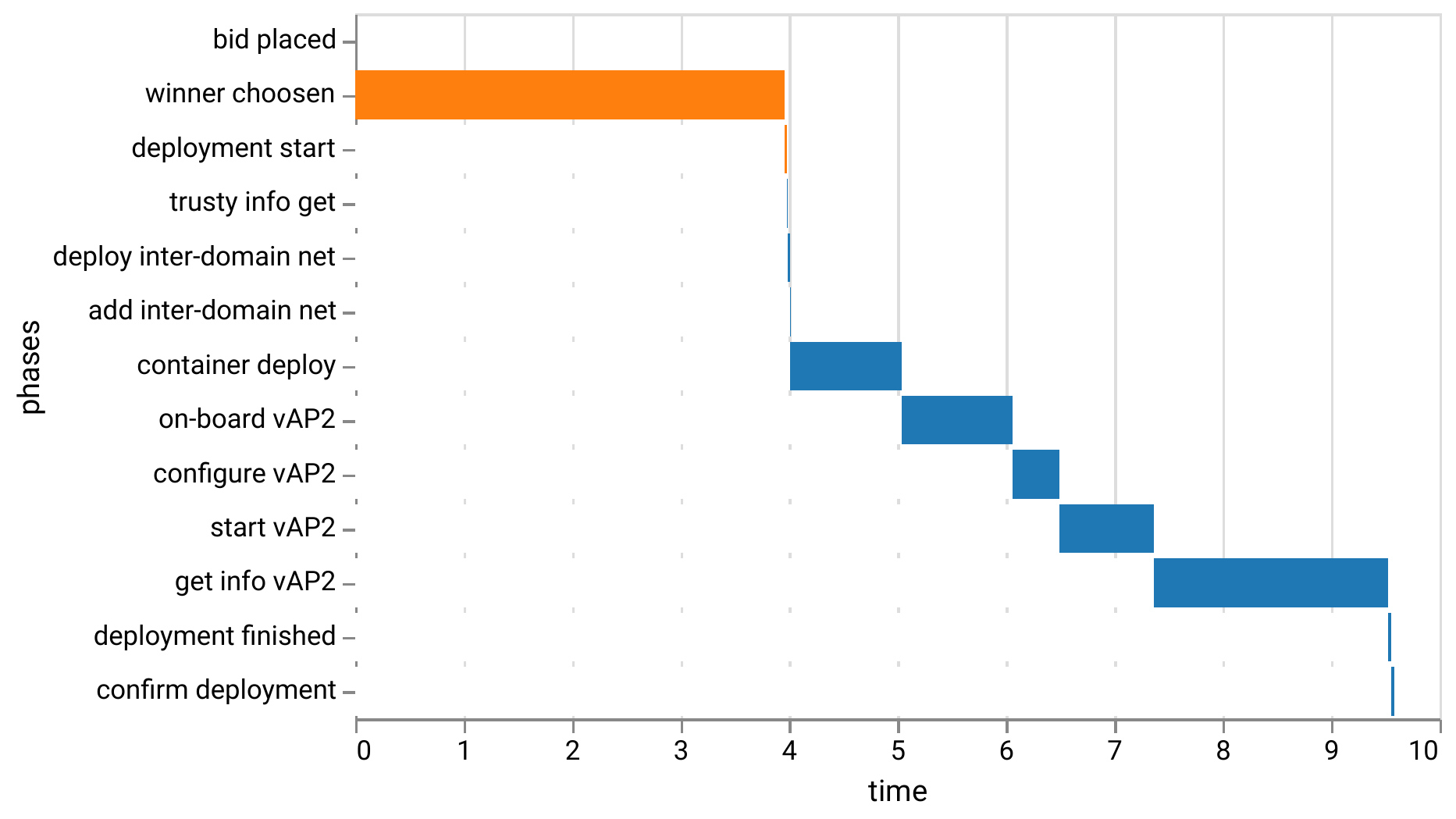}
		\vspace{-5 mm}
		
	\end{subfigure}
	
	\label{fig:PoA_fed}
	\caption{Federation using trusty communication - PoA consensus: (top) summarized phase times; (middle) consumer AD; (bottom) provider AD;}
	\vspace{-1.5 mm}
\end{figure}

Three graphs of the time it takes to finalize all the federation procedures are shown on Fig.~\ref{fig:PoA_fed}. In all graphs the time bars are colored:
\begin{itemize}
	\item orange - for all federation related procedures as described in Sec.~\ref{subsec:edge_federation}/\ref{subsec:dlt_application} and Fig.~\ref{fig:federation_sc}.
	\item blue - for all procedures that involve deployment of the Edge robotics service or part of it.
\end{itemize}

To that end, the top graph of Fig.~\ref{fig:PoA_fed} presents the accumulated times of the federation procedures in both consumer and provider domain.
The average federation time is 19.038 seconds - or the time it takes from the trigger at the consumer orchestrator to the robot connected to the \emph{vAP2}. 
The break-down in all phases that occur in the consumer domain is presented in the middle graph of Fig.~\ref{fig:PoA_fed}. 
It takes 12.97 seconds for the deployment of the \emph{vAP2} to be confirmed at the consumer domain (or phase "federation completed"). In other words, the consumer domain retrieves the BSSID of \emph{vAP2} in provider domain in 12.97 seconds. Then it takes around 6 seconds. for the \emph{Brain} to instruct the robot to discover \emph{vAP2}, disconnect from \emph{vAP1}, and connect to \emph{vAP2}.

The bottom graph of Fig.~\ref{fig:PoA_fed} breaks down all the phases in the provider domain, that occur within the previously mentioned 12.97 seconds. The negotiation or bidding process until the provider domain is elected as a winning provider takes 3.98 seconds. More specifically, it takes 3.98 seconds from the time that the provider domain receives the broadcast announcement (shown on Fig.~\ref{fig:federation_sc}) until the deployment starts. The establishment of the inter-domain link, on-boarding \& instantiation of the \emph{vAP2} takes additional 5.58 seconds. 

\begin{figure}
	\centering
	\vspace{-2.5 mm}
	\includegraphics[width=\columnwidth]{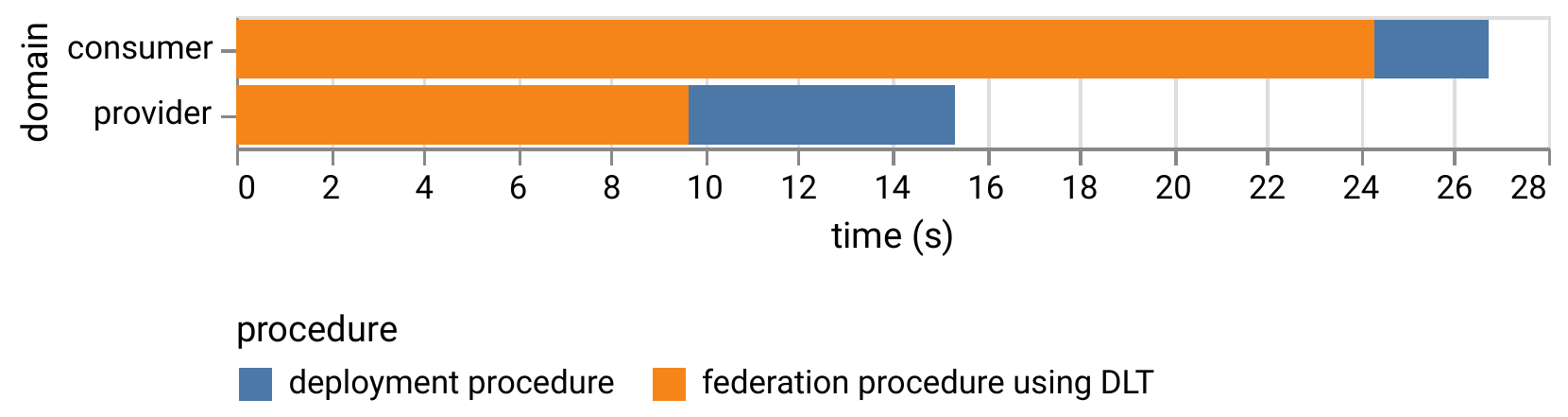}
	\caption{Federation using untrusty communication - PoW consensus: summarized times}
	\vspace{-6 mm}
	\label{fig:PoW_fed}
\end{figure}

The results of the PoW-based scenario and untrusty communication is shown in the Fig.~\ref{fig:PoW_fed}. The graph shows only the accumulated times for both domains. Compared to the PoA-based solution, it is clear that the PoW-based solution takes significantly more time to negotiate and complete the federation process using the Blockchain/DLT. Due to the PoW consensus mechanism the "federation completed" phase is completed within 24.3 seconds, nearly double the time of the PoA-based solution. 

\section{Conclusion}
\label{sec:conclusion}


This work showcases the application of DLT for federation of an Edge robotics service in a real scenario over an experimental test-bed. To the best of our knowledge, it is the first work that applies federation in an Edge robotics scenario. 

The results show that a complete federation process is concluded in around 19 seconds with lower level of security. In 42\% of the federation time, the consumer domain generates announcement, collects bids and chooses a winning provider domain. 
For higher security and anonymous negotiation among domains, the federation concludes in 28 seconds.  

A planned future work is to employ different DLTs (e.g., Hyperledger Fabric) with other consensus mechanisms and
apply various negotiation techniques 

\section*{Acknowledgments}
This work has been partially funded by the EU H2020 5GROWTH Project (grant no. 856709) and by the H2020 collaborative Europe/Taiwan research project 5G-DIVE (grant no. 859881).

\bibliographystyle{IEEEtran}
\bibliography{bibliography}

\begin{thebibliography}{10}
\providecommand{\url}[1]{#1}
\csname url@samestyle\endcsname
\providecommand{\newblock}{\relax}
\providecommand{\bibinfo}[2]{#2}
\providecommand{\BIBentrySTDinterwordspacing}{\spaceskip=0pt\relax}
\providecommand{\BIBentryALTinterwordstretchfactor}{4}
\providecommand{\BIBentryALTinterwordspacing}{\spaceskip=\fontdimen2\font plus
\BIBentryALTinterwordstretchfactor\fontdimen3\font minus
  \fontdimen4\font\relax}
\providecommand{\BIBforeignlanguage}[2]{{%
\expandafter\ifx\csname l@#1\endcsname\relax
\typeout{** WARNING: IEEEtran.bst: No hyphenation pattern has been}%
\typeout{** loaded for the language `#1'. Using the pattern for}%
\typeout{** the default language instead.}%
\else
\language=\csname l@#1\endcsname
\fi
#2}}
\providecommand{\BIBdecl}{\relax}
\BIBdecl

\bibitem{cloudrobotics}
B.~{Kehoe}, S.~{Patil}, P.~{Abbeel}, and K.~{Goldberg}, ``A survey of research
  on cloud robotics and automation,'' \emph{IEEE Transactions on Automation
  Science and Engineering}, vol.~12, no.~2, pp. 398--409, 2015.

\bibitem{so_5gt_orch_fed}
X.~Li~et al., ``Service orchestration and federation for verticals,'' in
  \emph{2018 IEEE Wireless Communications and Networking Conference Workshops
  (WCNCW)}, April 2018, pp. 260--265.

\bibitem{5gt_demo}
J.~B. et~al., ``Composing services in 5g-transformer,'' in \emph{Proceedings of
  the Twentieth ACM International Symposium on Mobile Ad Hoc Networking and
  Computing}, ser. Mobihoc ’19.\hskip 1em plus 0.5em minus 0.4em\relax New
  York, NY, USA: ACM, 2019, p. 407–408.

\bibitem{kiril_2}
K.~Antevski and C.~J. Bernardos, ``Federation of 5g services using distributed
  ledger technologies,'' \emph{Internet Technology Letters}.

\bibitem{bitcoin_nakamoto}
S.~Nakamoto, ``Bitcoin: A peer-to-peer electronic cash system,'' Manubot, Tech.
  Rep., 2019.

\bibitem{consensus2_survey}
L.~Bach, B.~Mihaljevic, and M.~Zagar, ``Comparative analysis of blockchain
  consensus algorithms,'' in \emph{2018 41st International Convention on
  Information and Communication Technology, Electronics and Microelectronics
  (MIPRO)}.\hskip 1em plus 0.5em minus 0.4em\relax IEEE, 2018, pp. 1545--1550.

\bibitem{ethereum_main}
G.~Wood \emph{et~al.}, ``Ethereum: A secure decentralised generalised
  transaction ledger,'' \emph{Ethereum project yellow paper}, vol. 151, no.
  2014, pp. 1--32, 2014.

\bibitem{edge_robotics_1}
``Cognitive computing and wireless communications on the edge for healthcare
  service robots,'' \emph{Computer Communications}, vol. 149, pp. 99 -- 106,
  2020.

\bibitem{edge_robotics_2}
S.~Dey and A.~Mukherjee, ``Robotic slam: A review from fog computing and mobile
  edge computing perspective.''\hskip 1em plus 0.5em minus 0.4em\relax New
  York, NY, USA: ACM, 2016.

\bibitem{edge_robotics_3}
N.~Tian, A.~K. Tanwani, J.~Chen, M.~Ma, R.~Zhang, B.~Huang, K.~Goldberg, and
  S.~Sojoudi, ``A fog robotic system for dynamic visual servoing,'' in
  \emph{2019 International Conference on Robotics and Automation (ICRA)}.\hskip
  1em plus 0.5em minus 0.4em\relax IEEE, 2019, pp. 1982--1988.

\bibitem{conext}
K.~A. et~al., ``Enhancing edge robotics through the use of context
  information.''\hskip 1em plus 0.5em minus 0.4em\relax New York, NY, USA: ACM,
  2018.

\bibitem{etsi.mec.017}
ETSI, ``{Mobile Edge Computing (MEC); Deployment of Mobile Edge Computing in an
  NFV environment},'' European Telecommunications Standards Institute (ETSI),
  Group Report (GR) 017 V1.1.1, 2 2018.

\bibitem{mec_in_nfv_kiril}
K.~A. et~al., ``On the integration of nfv and mec technologies: architecture
  analysis and benefits for edge robotics,'' \emph{Computer Networks}, vol.
  175, p. 107274, 2020.

\bibitem{jordi_vtmag}
J.~{Baranda}et al., ``Realizing the network service federation vision: Enabling
  automated multidomain orchestration of network services,'' \emph{IEEE
  Vehicular Technology Magazine}, vol.~15, no.~2, pp. 48--57, 2020.

\bibitem{5gt_infocom}
J.~B. et~al., ``Nfv service federation: enabling multi-provider ehealth
  emergency services,'' in \emph{Proceedings of the International Conference on
  Computer Communications (INFOCOM'20)}, July 2020.

\bibitem{reverse_auction}
S.~D. Jap, ``The impact of online reverse auction design on buyer–supplier
  relationships,'' \emph{Journal of Marketing}, vol.~71, no.~1, pp. 146--159,
  2007.

\bibitem{payment_ch_ethereum}
H.~G. et~al., ``An efficient micropayment channel on ethereum,'' in \emph{Data
  Privacy Management, Cryptocurrencies and Blockchain Technology}.\hskip 1em
  plus 0.5em minus 0.4em\relax Springer International Publishing, 2019, pp.
  211--218.

\end{thebibliography}

\end{document}